\documentclass[preprint,authoryear]{elsarticle}

\usepackage{aas_macros}

\usepackage[usenames]{color}

\newcommand{\VelaJr}{RX J0852.0-4622}
\newcommand{\model}[1]{\texttt{#1}}
\newcommand{\Jahr}{\mathrm{yr}}

\newcommand{\figref}[1]{Fig.\,\ref{#1}}
\newcommand{\tabref}[1]{Tab.\,\ref{#1}}
\newcommand{\eqref}[1]{Eq.\,(\ref{#1})}
\newcommand{\secref}[1]{Sect.\,\ref{#1}}

\newcommand{\Msol}{M_{\odot}}

\newcommand{\zehnh}[2]{{#1} \times 10^{#2}}
\newcommand{\inv}[1]{\frac{1}{#1}}

\newcommand{\cm}{\textrm{cm}}

\newcommand{\erg}{\textrm{erg}}

\newcommand{\iccm}{\textrm{cm}^{-3}}

\journal{High Energy Density Physics}

\begin{document}


\begin{frontmatter}

  \title{
    A method for computing synchrotron and inverse-Compton emission
    from hydrodynamic simulations of supernova remnants
  }

  \author[uv]{M.~Obergaulinger\corref{cor1}}
  \ead{martin.obergaulinger@uv.es}

  \author[uv]{J.M$^{\mathrm{a}}$~Chimeno Hern{\'a}ndez}
  \ead{jochiher@alumni.uv.es}

  \author[uv]{P.~Mimica}
  \ead{petar.mimica@uv.es}

  \author[uv]{M.{\'A}~Aloy Tor{\'a}s}
  \ead{miguel.a.aloy@uv.es}

  \author[srd,mpe]{A.~Iyudin}
  \ead{aiyudin@srd.sinp.msu.ru}

  \cortext[cor1]{Corresponding author}

  \address[uv]{Departament d{\'{}}Astronomia i Astrof{\'i}sica, Universitat de
    Val{\`e}ncia, \\ Edifici d{\'{}}Investigaci{\'o} Jeroni Munyoz, C/
    Dr.~Moliner, 50, E-46100 Burjassot (Val{\`e}ncia), Spain}

  \address[srd]{Extreme Universe Laboratory, Skobeltsyn Institute of Nuclear 
    Physics, Moscow State University by M.V. Lomonosov, \\ Leninskie Gory,
    119991 Moscow, Russian Federation}

  \address[mpe]{Max-Planck-Institut f{\"u}r Extraterrestrische Physik,
    Postfach 1312 D-85741 Garching, Bavaria, Germany}

  \begin{abstract}
    The observational signature of supernova remnants (SNRs) is very
    complex, in terms of both their geometrical shape and their
    spectral properties, dominated by non-thermal synchrotron and
    inverse-Compton scattering.  We propose a post-processing method
    to analyse the broad-band emission of SNRs based on
    three-dimensional hydrodynamical simulations.  From the
    hydrodynamical data, we estimate the distribution of non-thermal
    electrons accelerated at the shock wave and follow the subsequent
    evolution as they lose or gain energy by adiabatic expansion or
    compression and emit energy by radiation.  As a first test case,
    we use a simulation of a bipolar supernova expanding into a cloudy
    medium.  We find that our method qualitatively reproduces the main
    observational features of typical SNRs and produces fluxes that
    agree with observations to within a factor of a few.  allowing for
    further use in more extended sets of models.
  \end{abstract}


\end{frontmatter}


\section{Introduction}
\label{Sec:Intro}

Supernova remnants (SNRs) are characterised by electromagnetic
emission across a wide spectral range, which is generated by several
different emission mechanisms such as (thermal) bremsstrahlung,
synchrotron and inverse Compton scattering (IC), and the line emission
of many different chemical elements in various ionisation levels
\citep[see,
e.g.,][]{Gould__1965__PRL__High-EnergyPhotonsfromtheCompton-SynchrotronProcessintheCrabNebula,deJager_Harding__1992__apj__Theexpectedhigh-energytoultra-high-energygamma-rayspectrumoftheCrabNebula}.
Similar processes play an important role in many other astrophysical
objects
\citep[e.g.,][]{Zdziarski__1986__apj__OntheoriginoftheinfraredandX-raycontinuaofactivegalacticnuclei}.
Consequently, the spectra and lightcurves depend on many physical
processes, some of which can be--within a broad range of
uncertainties--inferred from observational data (explosion energy,
properties of the environment. Unfortunately, there are other key
processes which may be not fully understood such as the physics of
radiative shocks and the associated particle
acceleration. Furthermore, many SNRs show a complex geometry with
deviations from spherical symmetry that may be attributed to
combinations of asymmetries in the explosion, instabilities during the
propagation of the shock wave such as Rayleigh-Taylor, Vishniac (thin
shell) or unstable cooling \citep[see,
e.g.,][]{Chevalier_Imamura__1982__apj__Linearanalysisofanoscillatoryinstabilityofradiativeshockwaves},
inhomogeneities in the surrounding interstellar medium (ISM), or
magnetic fields.  For a review on observations of SNRs, we refer to
\cite{Reynolds__2008__araa__SupernovaRemnantsatHighEnergy}.  Among the
galactic remnants, SNR \VelaJr~(Vela Jr.) is a particularly
interesting case, with observations from radio to TeV energies
revealing a complex emission geometry
\citep{Duncan_Green__2000__aap__ThesupernovaremnantRXJ0852.0-4622_radiocharacteristicsandimplicationsforSNRstatistics,Slane_et_al__2001__apj__RXJ0852.0-4622:AnotherNonthermalShell-TypeSupernovaRemnant(G266.2-1.2),Aharonian_et_al__2007__apj__H.E.S.S.ObservationsoftheSupernovaRemnantRXJ0852.0-4622_Shell-TypeMorphologyandSpectrumofaWidelyExtendedVeryHighEnergyGamma-RaySource,Iyudin_et_al__2007__ESASP__MultiwavelengthAppearanceofVelaJr.:IsituptoExpectations,Tanaka_et_al__2011__apjl__Gamma-RayObservationsoftheSupernovaRemnantRXJ0852.0-4622withtheFermiLargeAreaTelescope,Kishishita_et_al__2013__aap__NonthermalemissionpropertiesofthenorthwesternrimofsupernovaremnantRXJ0852.0-4622}.

The complexity of the radiation processes and the hydrodynamics of the
SNR restrict a straightforward interpretation of observations and
require the use of increasingly complex models in order to understand
the physics of SNRs in general and of individual objects.  Depending
on the objective, models may focus on different effects, while making
simplification in other sectors of physics.  For instance,
\cite{Obergaulinger_et_al__2014__mnras__HydrodynamicsimulationsoftheinteractionofsupernovashockwaveswithaclumpyenvironmentthecaseoftheRXJ0852.0-4622(VelaJr)supernovaremnant}
performed a series of three-dimensional simulations of the expansion
of supernova blast waves into a clumpy environment \citep[for models
of simular settings, see][]{Orlando_et_al__2005__aap__CrushingofinterstellargascloudsinsupernovaremnantsITheroleofthermalconductionandradiativelosses,Orlando_et_al__2006__aap__CrushingofinterstellargascloudsinsupernovaremnantsIIX-rayemission}.  Paying attention
in particular to the case of Vela Jr., they concentrated their efforts
on an accurate modelling of the hydrodynamics of the interaction
between the shock wave and clouds in the ISM.  Their model for the
electromagnetic emission, on the other hand, was relatively limited
and accounted only for thermal bremsstrahlung, leaving out some of the
most important contributions to the emission coming from SNRs.

Our goal now is to remedy this limitation by modelling the non-thermal
emission of the simulated SNRs.  To this end, we propose a method for
post-processing the existing simulations.  We assume that particle
acceleration at the shock wave generates a population of high-energy
electrons that subsequently cool by synchrotron and inverse-Compton
radiation.  The advantage of our approach is that it can provide and
quick estimate of the non-thermal emission, but at a cost of several
simplifications w.r.t., e.g.~the spectra of the shock-accelerated
electrons, the seed photons for IC, and the magnetic field in the SNR.
Furthermore, we neglect ionisation and line emission.
\cite{Kishishita_et_al__2013__aap__NonthermalemissionpropertiesofthenorthwesternrimofsupernovaremnantRXJ0852.0-4622}
used similar methods to interpret observations of the Vela Jr.~SNR,
but coupled the emission model to spherically symmetric analytic
hydrodynamical models, whereas we will use self-consistent
hydrodynamical simulations to determine the evolution of the shock
wave.  Our method is, however, less accurate than, e.g.~the
simulations of
\cite{Lee_et_al__2013__apj__ACR-hydro-NEIModelofMulti-wavelengthEmissionfromtheVelaJr.SupernovaRemnant(SNRRXJ0852.0-4622)},
which couple hydrodynamics, non-equilibrium ionisation, non-linear
diffuse shock acceleration, and cosmic-ray production in a fully
self-consistent manner.  We also refer to the work of
\cite{Orlando_et_al__2011__aap__Effectsofnon-uniforminterstellarmagneticfieldonsynchrotronX-rayandinverse-Comptongamma-raymorphologyofsupernovaremnants}
for computations of the non-thermal emission in SNRs based on
multi-dimensional simulations.  We additionally note that our approach
is at an approximate level related to the relativistic emission
modelling of
\cite{Mimica_et_al__2004__aap__SyntheticX-raylightcurvesofBLLacsfromrelativistichydrodynamicsimulations}
and
\cite{Mimica_et_al__2010__mnras__Multiwavelengthafterglowlightcurvesfrommagnetizedgamma-rayburstflows}.
By virtue of its simplicity, the method lends itself easily to an
investigation of the impact of variation of the input physics, e.g.~of
the spectra of accelerated electrons or the seed photons for the IC
process.

We will begin the presentation in this article with a brief recap of
the hydrodynamical simulations of
\cite{Obergaulinger_et_al__2014__mnras__HydrodynamicsimulationsoftheinteractionofsupernovashockwaveswithaclumpyenvironmentthecaseoftheRXJ0852.0-4622(VelaJr)supernovaremnant}
and an outline of our emission model in \secref{Sek:PhysNum}, then
present results for the non-thermal radiation emitted by one of the
models (\secref{Sek:Res}), before summarising the main results and
drawing further conclusions in \secref{Sek:Sum}.

\section{Physical ingredients and numerical method}
\label{Sek:PhysNum}

Following the implementation in the SPEV code outlined in
\cite{Mimica_et_al__2009__apj__SpectralEvolutionofSuperluminalComponentsinParsec-ScaleJets},
we model the evolution of a population of non-thermal electrons
accelerated by the shock wave and their subsequent emission of
synchrotron and IC radiation using the post-processing algorithm
\emph{SPEVita} based on three-dimensional hydrodynamic simulations.
Such a two-step approach requires that the radiative energy losses do
not significantly alter the structure and evolution of the remnant.
This condition is satisfied in our cases because the total amount of
energy carried away by photons is small w.r.t.~the total kinetic and
internal energy of the remnant.

We furthermore work in the limit of low optical depth of the gas in
the SNR.  This assumption is justified by the low gas density.  If we
estimate the optical depth of the SNR using the Thomson scattering
cross section, $\sigma_{\mathrm{Th}} = 0.66 \times 10^{-24} \, \cm^2$,
we find that even the densest clouds in our simulations (densities up
to $1000 \, \iccm$, size up to 10 pc) have optical depths of less than
0.01, and most of the gas is much more optically thin.  This fact
allows us to directly obtain the radiation arriving at an observer
location from the emissivity at the source location instead of solving
the much more complex equations of radiative transfer.  Furthermore,
we neglect synchrotron self-absorption since it is unimportant in the
frequencies considered here (above the optical band).

\paragraph{Hydrodynamical models}

From the simulations of
\cite{Obergaulinger_et_al__2014__mnras__HydrodynamicsimulationsoftheinteractionofsupernovashockwaveswithaclumpyenvironmentthecaseoftheRXJ0852.0-4622(VelaJr)supernovaremnant},
we select a model (model \model{S25A}) in which a bipolar supernova
explosion ejects a mass of $M_{\mathrm{SN}} = 6 \Msol$ with a total
explosion energy of $E_{\mathrm{SN}} = \zehnh{6.7}{51} \, \erg$ into
an ISM of particle density $n_{\mathrm{ISM}} = 0.25 \, \iccm$ and
temperature $T_{\mathrm{ISM}} = 10 \ \mathrm{K}$.  The ISM contains
four large high-density clouds placed in the NW, N, SE and S
directions from the the centre of the explosion at positions where
X-ray bright features are suggestive of an interaction between the
shock wave and overdense structures in the ISM.

The expanding shock wave roughly maintains its initial bipolar shape
with an interior consisting of a hot, tenuous gas with very little
substructure.  This changes once, after a time of about $t \sim 700 \,
\Jahr$, the interaction between the shock wave and the clouds channels
the expanding gas between the gas clouds and enhances the mixing of
post-shock fluid elements.  On the time scales under consideration
here, i.e.~up to 1500 years after the explosion, the clouds are not
disrupted by the shock, but experience considerable deformation and,
most importantly, heating of the shocked surfaces, which, as a
combination of high temperature and high density, show up as prominent
emitters of thermal radiation.

\paragraph{Non-thermal emission}

The passage of the shock wave across a fluid element generates a
population of relativistic non-thermal electrons.  Without modelling
the acceleration process in detail, we assume that the
$0^{\mathrm{th}}$ moment of the distribution function of the
electrons, $n^0 (\gamma)$, i.e.~the number density of particles per
unit Lorentz factor, $\gamma$, follows a power-law distribution,
\begin{equation}
  \label{Gl:acc-powerlaw}
  n^0 ( \gamma ) = n^0_0 
  \left( \frac{ \gamma} {\gamma_{\mathrm{min}}} \right) ^ { -q } \ 
  \mathrm{for} \  \gamma_{\mathrm{min}} \le \gamma \le \gamma_{\mathrm{max}}.
\end{equation}
For further reference, we note that the Lorentz factor is related to
the particle momentum and energy via the electron mass, $m_e$, as $p =
\gamma m_e v$ ($p = \gamma m_e c$ for the case of ultrarelativistic
electrons considered in the following) and $e = \gamma m_e c^2$,
respectively.  Aside from the power-law index $q$, there are three
{\em free} parameters that specify the electron energy distribution,
namely, the minimum and maximum Lorentz factors,
$\gamma_{\mathrm{min}}$ and $\gamma_{\mathrm{max}}$, and the
normalisation $n_0^0$.  The following conditions allow us to fix these
three parameters:
\begin{enumerate}
\item We first estimate the value of the stochastic magnetic field
  energy density generated at shocks assuming that it is a fraction,
  $\epsilon_b$, of the thermal energy density, $u_S$ (provided by our
  hydrodynamic models), i.e., $B=\sqrt{8\pi \epsilon_b u_S}$.
\item Following
  \cite{Reynolds__2008__araa__SupernovaRemnantsatHighEnergy}, we
  relate the cut-off energy of the electron distribution to the
  magnetic field strength at the site of acceleration (computed in
  point 1, above), $E_{\mathrm{max}} = 100 \, \mathrm{TeV} \
  \alpha_{\mathrm{acc}} \left( \frac{B}{1 \, \mu \mathrm{G}} \right)
  ^{-1/2}$, and thus, we shall specify the value of the parameter
  $\alpha_{\rm acc}$.
\item Given $\gamma_{\mathrm{max}}$, the normalisation $n^0_0$ and the
  minimum Lorentz factor are direct functions of the efficiency of the
  acceleration process, i.e.~the fraction of electrons accelerated in
  the shock wave and the fraction of total energy they carry,
  $\epsilon_n$ and $\epsilon_e$, respectively.
\end{enumerate}
In practise, we ignore all electrons below $\gamma_{\rm min; emi} =
10$ when computing synchrotron and IC emission to be consistent with
the approximations we will employ in their respective emissivities.

We consider only a single episode of particle acceleration.  Upon
passage of the shock wave across one of our tracer particles, we
define a momentum grid of $n_p$ zones distributed logarithmically
spanning the range $[p_{\mathrm{min}},\, p_{\mathrm{max}}] =
[\gamma_{\mathrm{min}},\, \gamma_{\mathrm{max}}]\times m_e c$ and
initialise $n^0 (p_i), i = 1,..., n_p$ according to
\eqref{Gl:acc-powerlaw}.

Afterwards, the electrons suffer radiative losses and gain or lose
energy due to adiabatic compression or expansion, respectively.  In
the evolution equation for the $0^{\mathrm{th}}$ moment,
\begin{equation}
  \label{Gl:n0-evo}
  D_t \ln n^0 
  + \left( - \frac{p}{3} \Theta + \mathcal{B} \right)
  \partial_{p} \ln n^0
  =
  - \frac{2}{3} \Theta - \partial_{p} \mathcal{B},
\end{equation}
these effects are accounted for by the expansion coefficient 
\begin{equation}
  \label{Gl:Theta-def}
  \Theta =  - D_t \ln \rho
\end{equation}
 and the emission coefficient 
 \begin{equation}
   \label{Gl:B-def}
   \mathcal{B} = - \frac{4\sigma_{\mathrm{T}} (u_B +
     u_{\mathrm{ph}})}
   {3 m_e^2 c^2} p^2.
\end{equation}
Here, $D$ denotes the Lagrangian time derivative, $\rho$ is the gas
density, and $\sigma_{\mathrm{T}}$ is the Thomson cross section.  The
emission coefficient is the sum of a synchrotron contribution,
proportional to the magnetic energy density $u_B = \vec B^2 / 2$, and
an IC contribution, proportional to the energy density of the
background photon field, $u_{\mathrm{ph}}$.  $u_B$ and $u_{\rm ph}$
are, besides the efficiencies determining the initial distribution of
non-thermal particles, the most important free parameters of our
analysis.

The formal solution of \eqref{Gl:n0-evo}, given by
\cite{Mimica_et_al__2009__apj__SpectralEvolutionofSuperluminalComponentsinParsec-ScaleJets},
lends itself straightforwardly to a discretisation on a time-dependent
grid in momentum space.  Given the momentum of an electron $p ( t_0)$
at time $t_0$ and setting
\begin{equation}
  \label{Gl:ka-def}
  \frac{\rho (t_1)}{\rho (t_0)} = \exp{(3 k_{\mathrm{a}} \delta t)}
\end{equation}
and
\begin{equation}
  \label{Gl:ke-def}
  \mathcal{B} = - k_{\mathrm{e}} \delta t,
\end{equation}
we find the Lorentz factor of electrons at time $t_1 = t_0 + \delta t$
as a function of their Lorentz factor at time $t_0$
\begin{equation}
  \label{Gl:p-evol}
  \gamma ( t_1 )
  = 
  \gamma ( t_0 )
  \frac{ k_{\mathrm{a}} \exp{(k_{\mathrm{a}} \delta t) }}
  {k_{\mathrm{a}} + p(t_0)(\exp{(k_{\mathrm{a}} \delta t) -1}) },
\end{equation}
and the $0^{\mathrm{th}}$ moment as
\begin{equation}
  \label{Gl:n0-evol}
  n^0 ( t_1 ) = 
  n^0 ( t_0 ) 
  \left[ 
    \exp{(k_{\mathrm{a}} \delta t) }
    \left( 
      1 + \gamma (t_0) \frac{k_{\mathrm{a}}}{k_{\mathrm{e}}}
      ( \exp{(k_{\mathrm{a}} \delta t )} - 1  )
    \right)
  \right]^2.
\end{equation}
Solving Eqns.~\ref{Gl:p-evol} and \ref{Gl:n0-evol} for each tracer
particle that already passed by the shock wave, we obtain the time
evolution of the distribution of non-thermal particles at discrete
momentum values $p_i (t)$.  Between these interface values, we
approximate the function $n^0 ( \gamma)$ by piecewise power laws of
index $q$, i.e.
\begin{equation}
  \label{Gl:ppl-n}
  n ^0 ( \gamma) = n^0_i 
  \left( \frac{\gamma}{\gamma_i}\right)^{-q}
  \ \mathrm{for} \
  \gamma_i \le \gamma < \gamma_{i+1}.
\end{equation}

This procedure does not directly yield the spectral distribution of
the emitted radiation, which we compute in an additional step using
the expressions summarised by
\cite{Boettcher_Reimer__2012____RelativisticJetsfromActiveGalacticNuclei}.
The synchrotron emission at frequency $\nu$ of non-thermal electrons
is described by the emissivity coefficient, i.e.~the energy emitted
per unit volume, unit frequency interval, and unit time 
\begin{equation}
  \label{Gl:synchro-emi}
  j^{\mathrm{syn}} ( \nu ) 
  = \inv{4\pi} 
  \int \mathrm{d} \gamma \, n ( \gamma ) P_{\nu} ( \gamma ).
\end{equation}
The radiative output of the distribution of electrons can be computed
by integrating the product of the single-particle emission function,
\begin{equation}
  \label{Gl:synchro-P}
  P_{\nu} ( \gamma ) = 
  \frac{ 32 c}{9 \Gamma(4/3)}
  \left( \frac{q_e^2}{m_e c^2}\right)^2 
  u_B \gamma ^2 
  \frac{\nu^{1/3}}{\nu_{\mathrm{c}}^{4/3}}
  \exp{(- \nu / \nu_{\mathrm{c}})},
\end{equation}
and the power-law distribution \eqref{Gl:ppl-n} over Lorentz factor.
The critical frequency $\nu_c$ is given by $\nu_{\mathrm{c}} = \frac{3
  q_e B}{2 m_e c^2} \gamma^2$ ($q_e$ is the electron charge).  For IC,
we restrict ourselves to the simple case of a mono-energetic seed
field of photons of frequency $\nu_0$, leading to an emission
coefficient for the power-law spectrum of electrons with momentum $p_i
\le \gamma m_e c < p_{i+1}$ approximately given by
\begin{eqnarray}
  \label{Gl:IC-emi}
  j_i^{\mathrm{IC}} (\nu) =
  C \left( \frac{\nu}{\nu_{\mathrm{c}}}\right)^2
  \left\{ 
    \left[
      \mathrm{max} \left( \gamma_{i}, \frac{h \nu}{m_e c^2},
      \sqrt{\frac{\nu}{2\nu_{\mathrm{c}}}} \right)
    \right]^{-q-3}
    - \gamma_{i+1}^{-q-3}
  \right\}
\end{eqnarray}
if $\gamma_{i+1} \ge \mathrm{max}\left( \frac{h \nu}{m_e c^2},
  \sqrt{\frac{\nu}{2\nu_0}} \right)$ and $j^{\mathrm{IC}}_{\nu} = 0$
otherwise; the normalisation is given in terms of the number density
of seed photons, $n_{\mathrm{ph,0}}$, by $C = \frac{h c
  \sigma_{\mathrm{T}} n^0_i n_{\mathrm{ph,0}}}{8 \pi (q+3)
  \gamma_i^{-q}}$.

For each tracer particle and its non-thermal electron distribution, we
evaluate the emission coefficients at discrete frequencies between
infrared and TeV energies.  We determine emission maps, i.e.~the total
spectral emissivity $J_{\nu} (x,y; T)$ arriving at the position of an
observer at distance $D$ along the $z-$axis from gas at position
$(x,y)$ on the celestial plane by integrating the emissivity along the
trajectory of a ray of light at time $T$,
\begin{equation}
  \label{Gl:obs-int}
  J_{\nu}^{\mathrm{syn; IC}} ( x,y; T) =
  \int_{-\infty}^{D} \mathrm{d} z \, 
  j_{\nu}^{\mathrm{syn; IC}} ( x,y,z; t = T - \inv{c} (D- z)).
\end{equation}
For the spatial integration, we take into account the expansion of the
gas by assuming that each tracer particle represents a uniform
distribution of electrons centred at its position and with a volume $V
(t) = V(t_0) \frac{\rho(t)}{\rho(t_0)}$, where $V (t_0)$ denotes the
volume assigned to the particle at the start of the simulation.
Finally, an integration of $J_{\nu} (x,y;T)$ over $x$ and $y$ yields
the total, spatially unresolved emission of the gas.

\section{Results}
\label{Sek:Res}

\begin{table}
  \centering
  \begin{tabular}{l |  l l l l | l} 
    \hline
    Model &$ \alpha_{\mathrm{acc}} $ & $\epsilon_e $ & $\epsilon_n$ &
    $  \epsilon_b $ & Linestyle (Colour)
    \\ 
    \hline
    \hline
    Reference & 0.1 & 0.3 & 0.14 & $10^{-3}$ & Solid (black)  \\
    \hline
    1 & 0.01 & 0.3 & 0.14 & $10^{-3}$ & Triple dot-dash (blue) \\
    2 & 1 & 0.3 & 0.14 & $10^{-3}$ & Dashed (blue) \\ 
    \hline
    3 & 0.1 & 0.09 & 0.14 & $10^{-3}$ & Triple dot-dash (green)   \\
    4 & 0.1 & 0.45 & 0.14 & $10^{-3}$ & Dashed (green) \\
    \hline
    5 & 0.1 & 0.3 & 0.042 & $10^{-3}$ & Triple dot-dash (red)  \\
    6 & 0.1 & 0.3 & 0.21 & $10^{-3}$ & Dashed (red) \\
    \hline
    7 & 0.1 & 0.3 & 0.14 & $10^{-4}$ & Triple dot-dash (orange)  \\
    8 & 0.1 & 0.3 & 0.14 & $10^{-2}$ & Dashed (orange) \\ 
    \hline
  \end{tabular}
  \caption{
    List of our models.  For each model (name or number in the first
    column), we list the parameter $\alpha_{\mathrm{acc}}$ governing
    the acceleration efficiency,  the normalisations of the energy
    and number densities of the non-thermal electrons
    ($\epsilon_e$ and $\epsilon_n$), and the magnetisation parameter, $\epsilon_b$.  The
    last column gives the linestyle and colour used for the spectrum
    of the corresponding model in \figref{Fig:Parsvars}.
  }
  \label{Tab:models}
\end{table}

Keeping the hydrodynamical model fixed, we computed several model
light curves, spectra, and emission maps.  The goal of this series of
models is not first and foremost to provide a good fit to the
observations of a particular SNR, but to assess in principle the
viability of our model to reproduce the most important features of the
non-thermal emission from SNRs and to explore the dependence of the
results on the input physics and the free parameters governing the
evolution of the non-thermal electrons and their emission, viz.~the
efficiency of particle acceleration, the magnetic field in the SNR,
and the background photon field acting as seed for IC scattering.  To
facilitate our goal, we will qualitatively compare the models to the
observations of the Vela Jr.~SNR in radio
\citep{Duncan_Green__2000__aap__ThesupernovaremnantRXJ0852.0-4622_radiocharacteristicsandimplicationsforSNRstatistics},
X-ray \citep[\emph{ASCA} GIS,
][]{Aharonian_et_al__2007__apj__H.E.S.S.ObservationsoftheSupernovaRemnantRXJ0852.0-4622_Shell-TypeMorphologyandSpectrumofaWidelyExtendedVeryHighEnergyGamma-RaySource},
\emph{Fermi}-LAT
\citep{Tanaka_et_al__2011__apjl__Gamma-RayObservationsoftheSupernovaRemnantRXJ0852.0-4622withtheFermiLargeAreaTelescope},
and \emph{HESS}
\citep{Aharonian_et_al__2007__apj__H.E.S.S.ObservationsoftheSupernovaRemnantRXJ0852.0-4622_Shell-TypeMorphologyandSpectrumofaWidelyExtendedVeryHighEnergyGamma-RaySource}.
We defer a more rigorous comparison with observational data to an
upcoming study.

We investigate how the emission depends on four parameters, viz.~
$\alpha_{\mathrm{acc}}$, $\epsilon_e$, $\epsilon_n$, and $\epsilon_b$.
For a reference model, we set $\alpha_{\mathrm{acc}} = 0.1$,
$\epsilon_e = 0.3$, $\epsilon_n = 0.14$, and $\epsilon_b = 10^{-3}$.
We compare the results to four pairs of models, where we increase and
decrease each of the four parameters individually (see
\tabref{Tab:models} for a list of models).  In all cases, the
power-law index of the non-thermal electrons was set to $q = 2.2$.

The synchrotron emission then follows directly from the settings for
the magnetic field.  We represent the seed for IC scattering by a
simple model for the cosmic microwave background (CMB), viz.~a field of
photons of temperature $T_{\mathrm{IC};0} = 2.7 \ \mathrm{K}$ with a uniform
energy density of $u_{\mathrm{IC;0}} = 1 \,
\mathrm{eV}/\mathrm{cm}^3$.

\begin{figure}
  \centering
  \includegraphics[width=\textwidth]{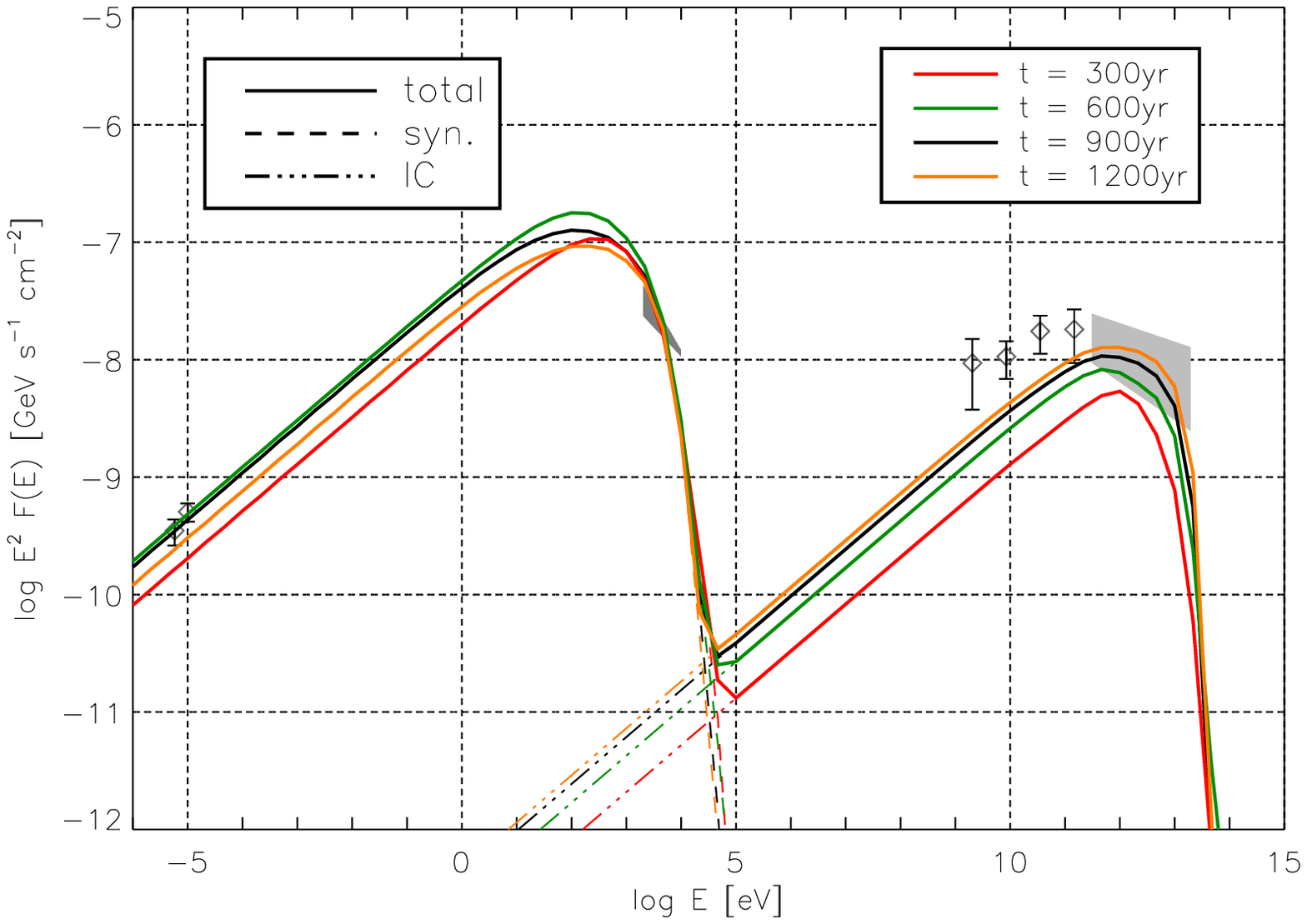}
  \caption{
    Broadband SED of our reference model for different times, distinguished
    by line colour.  The total emission and the synchrotron and IC
    contributions are displayed by solid, dashed, and
    dash-dot-dot-dotted lines, respectively.  For comparison, we
    include observations for the Vela Jr.~SNR in radio \citep[data
    points to the
    left;][]{Duncan_Green__2000__aap__ThesupernovaremnantRXJ0852.0-4622_radiocharacteristicsandimplicationsforSNRstatistics},
    X-ray \citep[\emph{ASCA} GIS; shaded region around 2 keV;
    ][]{Aharonian_et_al__2007__apj__H.E.S.S.ObservationsoftheSupernovaRemnantRXJ0852.0-4622_Shell-TypeMorphologyandSpectrumofaWidelyExtendedVeryHighEnergyGamma-RaySource},
    \emph{Fermi}-LAT \citep[data points in the GeV
    range;][]{Tanaka_et_al__2011__apjl__Gamma-RayObservationsoftheSupernovaRemnantRXJ0852.0-4622withtheFermiLargeAreaTelescope},
    and \emph{HESS} \citep[shaded region to the 
    right;][]{Aharonian_et_al__2007__apj__H.E.S.S.ObservationsoftheSupernovaRemnantRXJ0852.0-4622_Shell-TypeMorphologyandSpectrumofaWidelyExtendedVeryHighEnergyGamma-RaySource}.  
  }
  \label{Fig:refmod-spec}
\end{figure}

We show the spectral energy distribution of the radiation emitted by
the shock-accelerated electrons at four different times in
\figref{Fig:refmod-spec}.  Synchrotron radiation dominates the total
radiative output at low energies with a broad peak in the range of UV
to X-ray energies.  The spectral indices and the cutoff frequencies of
both components of the emission show little evolution with time, and
consequently the spectra at different times have rather similar
shapes.  We find, on the other hand, considerable differences in the
luminosities with time.  The synchrotron radiation increases from $t =
300 \, \Jahr$ and is greatest at $t = 600 \, \Jahr$, before it
gradually decreases again.  For times $t = 600 \, \Jahr$ and $t = 900
\, \Jahr$, our results agree very well with the radio observations.
For all times, the ASCA data lie in the steeply declining part of the
synchrotron spectrum.  The IC contribution is, in contrast to
synchrotron, increasing steadily throughout the entire simulation
time, though the brightening seems to slow down as we go from $t = 900
\, \Jahr$ to $t = 1200 \, \Jahr$.  During this period, our results
roughly match the observational data points.  The main discrepancy is
that our models do not reproduce the rather broad shape of the
observed IC maximum.  Assuming a thermal distribution of CMB seed
photons rather than the monochromatic spectrum we used might bring our
results closer to the observations.

\begin{figure}
  \centering
  \includegraphics[width=0.48\textwidth]{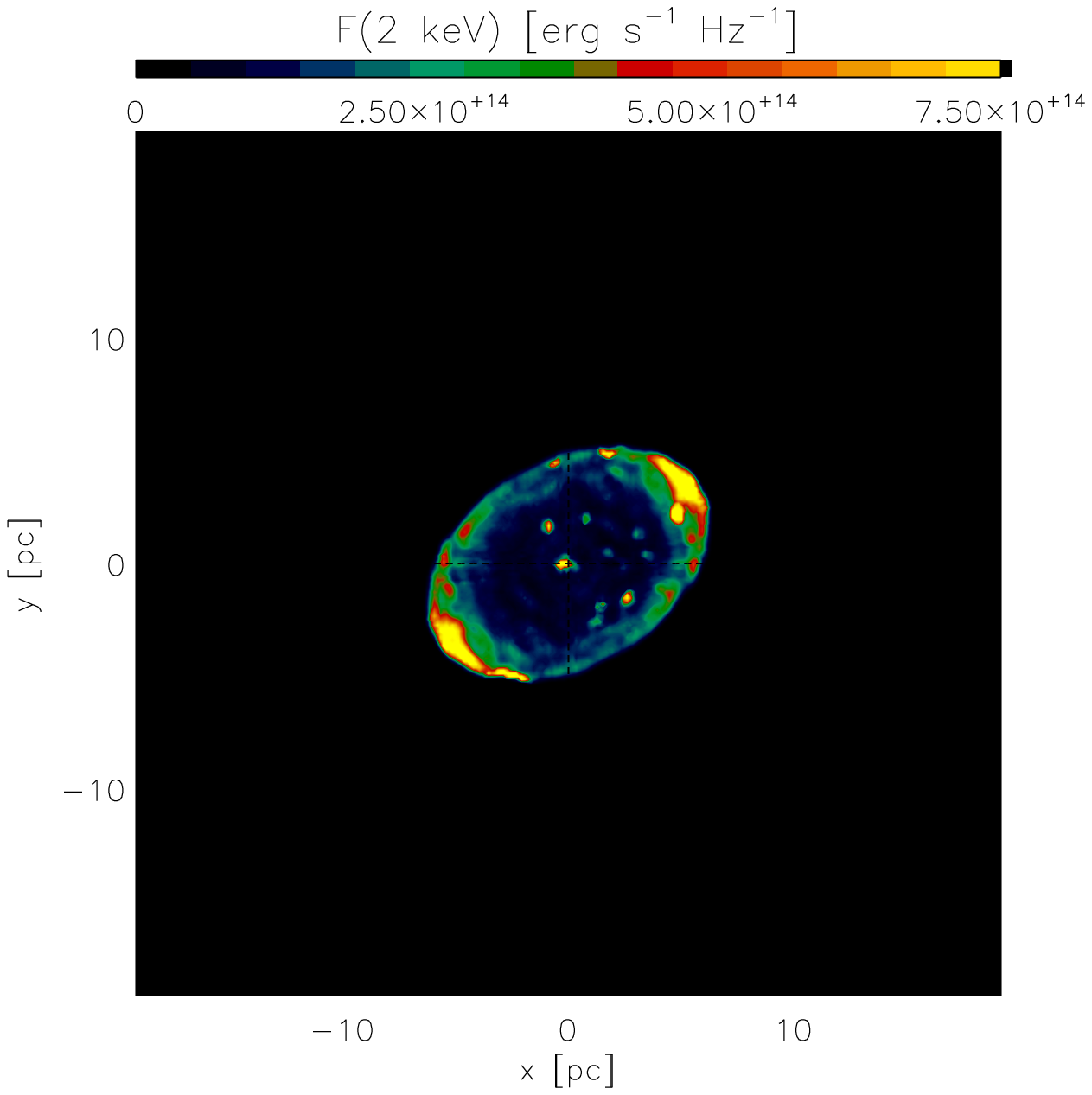}
  \includegraphics[width=0.48\textwidth]{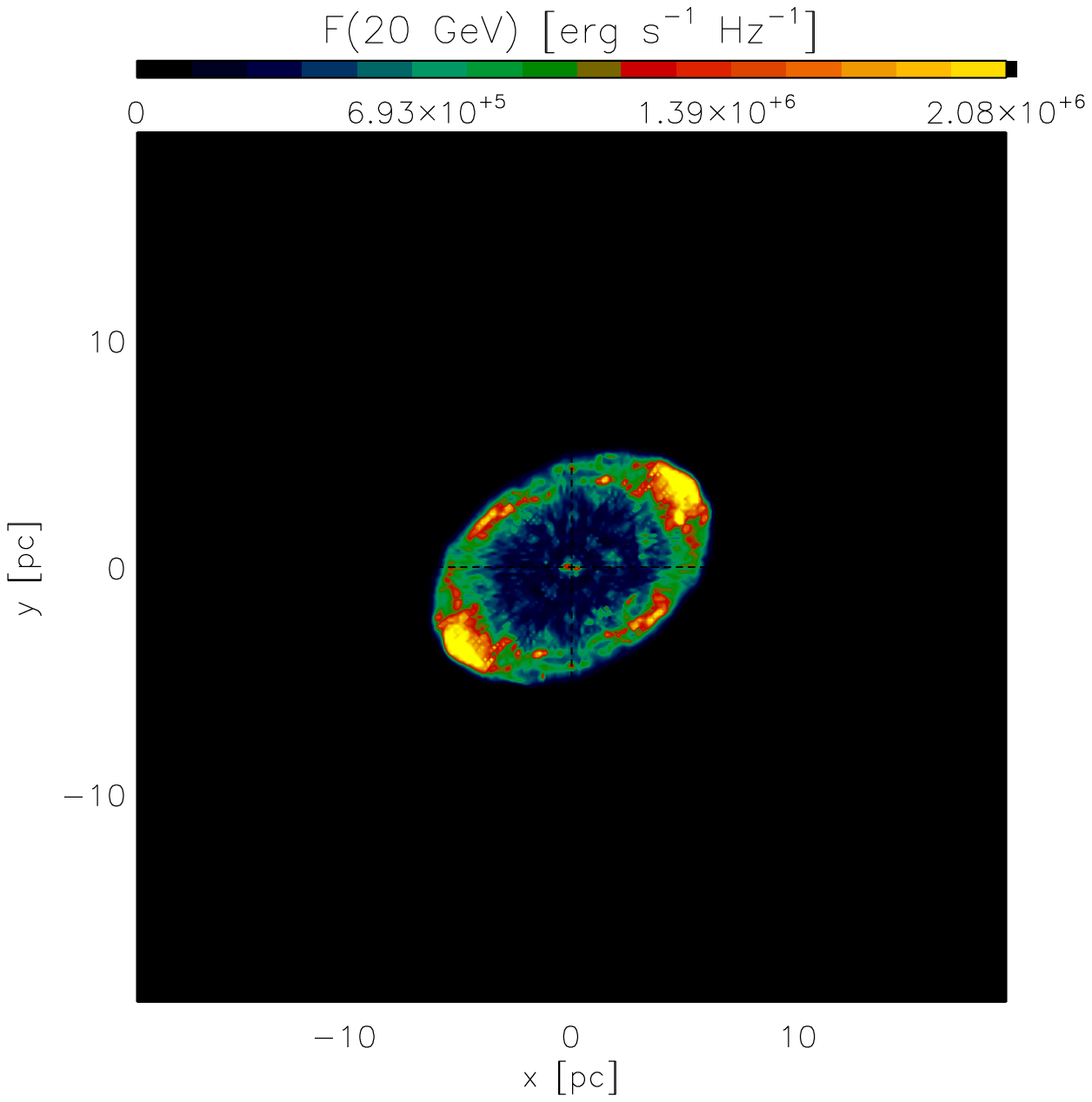}
  \includegraphics[width=0.48\textwidth]{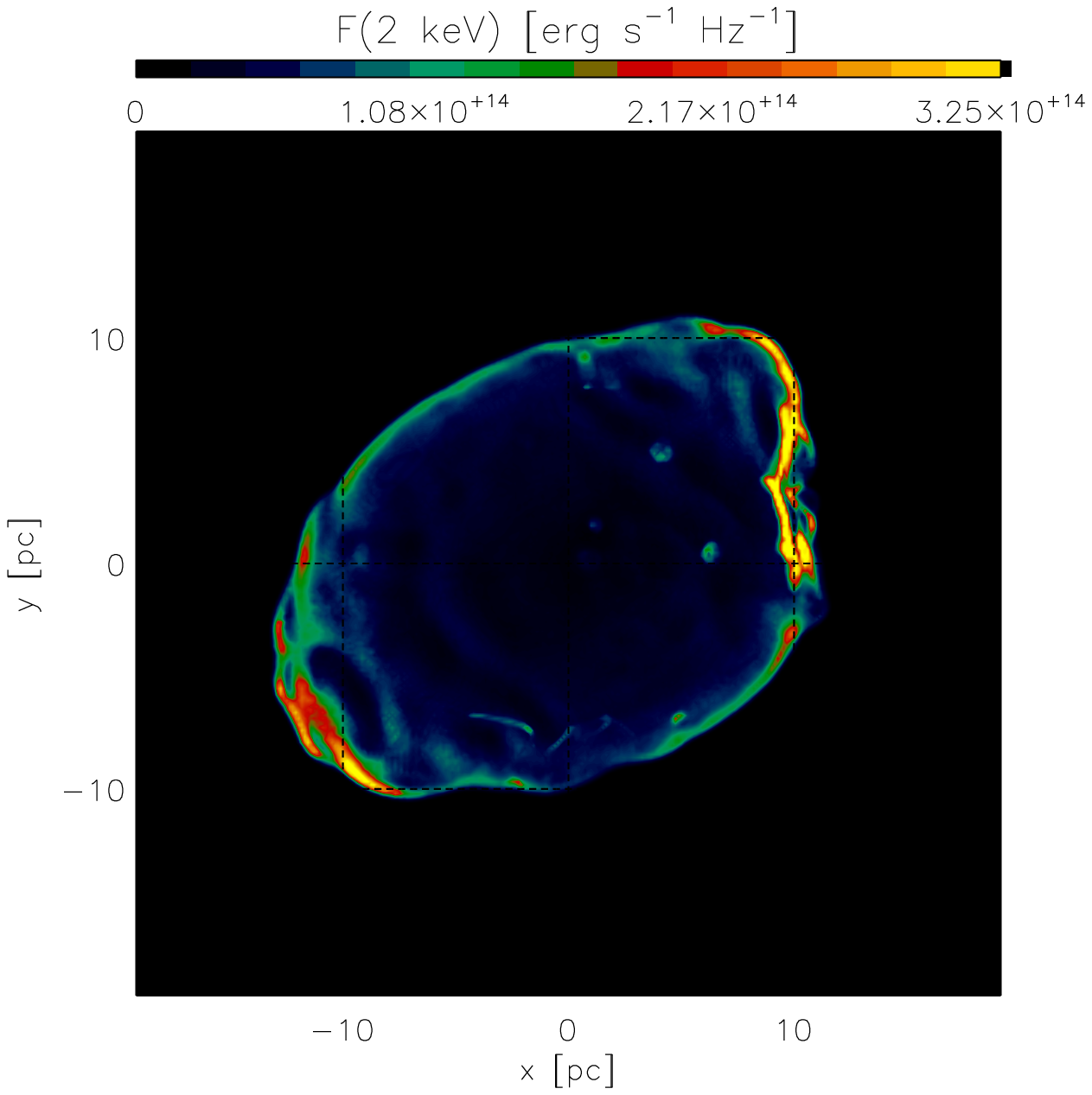}
  \includegraphics[width=0.48\textwidth]{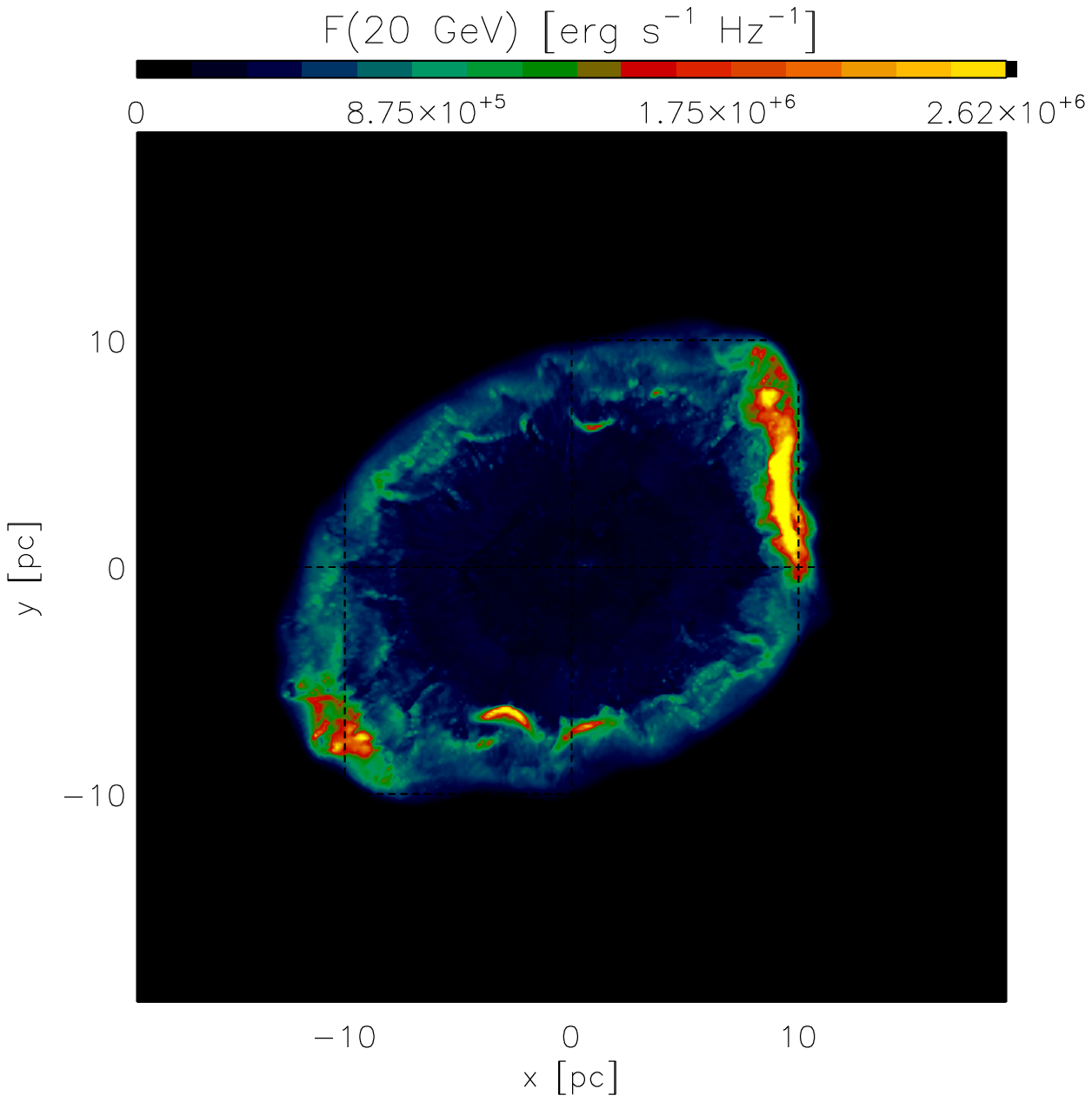}
  \caption{
    Emission maps in the X-ray (\textit{left}) and GeV
    (\textit{right}) bands at
    $t = 300 \, \Jahr$ (\textit{top}) and $t = 900 \, \Jahr$ (\textit{bottom}).
  }
  \label{Fig:refmod-maps}
\end{figure}

We investigate the evolving appearance of the SNR in the series of
emission maps presented in \figref{Fig:refmod-maps}.  Before hitting
the four major clouds (\textit{top} panels), the expanding shock wave
maintains its ellipsoidal shape, and consequently the observational
display of the SNR at low and high photon energies is dominated by the
limb-brightened shock wave.  The most prominent features are located
at the shock wave along the semi-major axis of the ejecta.  The
non-thermal particles cool rather rapidly, and, hence, the interior
regions of the SNR behind the shock wave remain dark.  A few bright
spots in the interior of the SNR indicate the positions of small
clouds already hit by the shock wave.  At a later time
(\textit{bottom} panels), the shock wave has already crossed the four
main clouds, which now show up as additional emitters, in particular
in the X-ray band.  This enhanced emission reflects the rather high
particle density and pressure at the shock-heated cloud surface as
well as the reduced velocities in the flow hitting the cloud, where
the fluid remains trapped for a fairly long time, thus reducing the
adiabatic cooling of the non-thermal electrons.

\begin{figure}
  \centering
  \includegraphics[width=\textwidth]{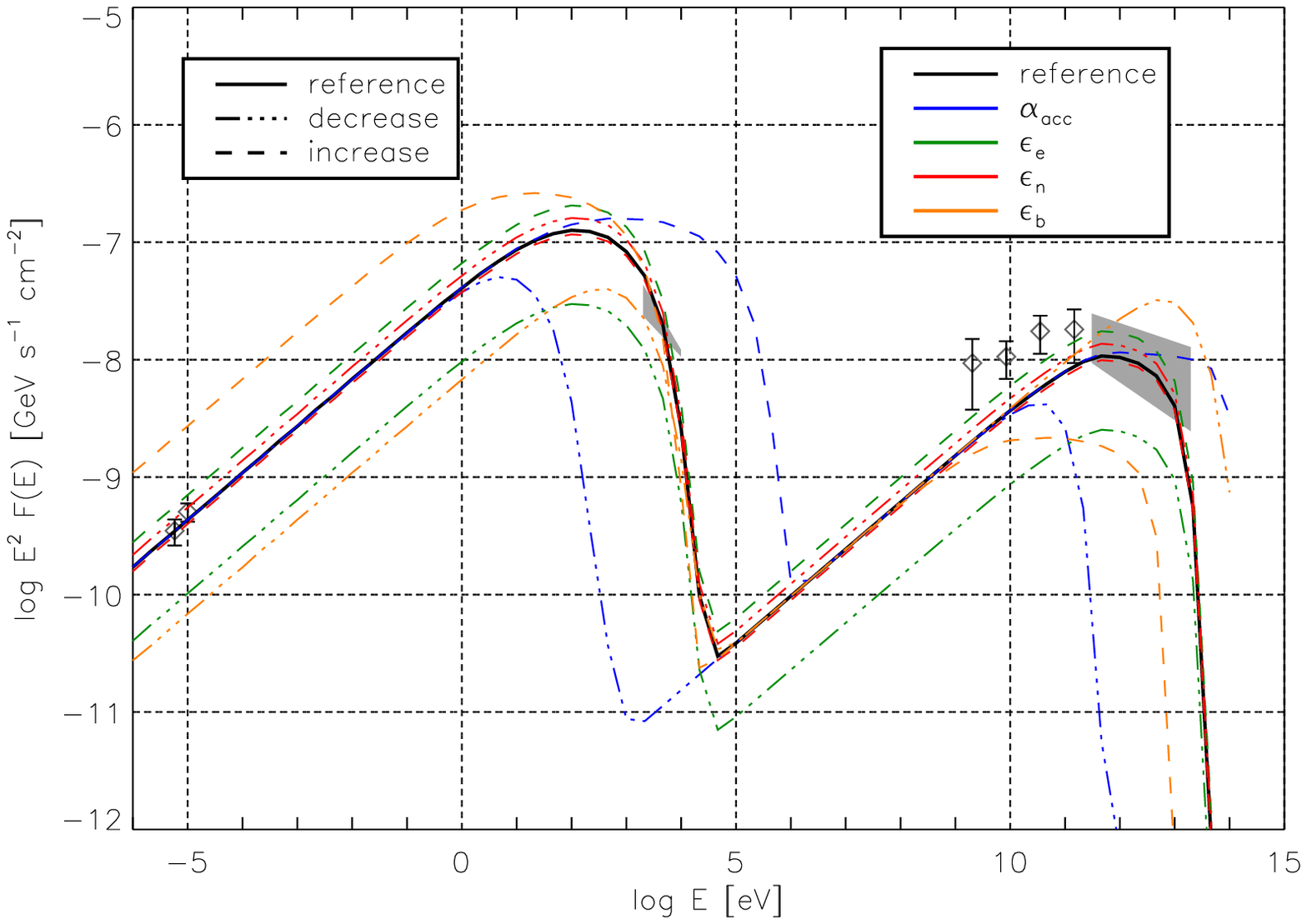}
  \caption{
    Comparison of the models with variations of the input parameters
    to the reference model (solid black line).  The lines show the total emission at $t
    = 900 \,  \Jahr$ for
    each model.  For each parameter, we computed a model with
    increased (dash-triple-dotted line) and a model with decreased
    (dashed line)
    value.  Line colours distinguish between the four parameters (see
    also \tabref{Tab:models}).
  }
  \label{Fig:Parsvars}
\end{figure}

We compare the results of the eight additional models in which we vary
one input parameter at a time in \figref{Fig:Parsvars}.  Lines of the
same colour belong to the same pair of models, in which one parameter
is either increased (dashed lines) or decreased (dash-triple-dotted
lines) w.r.t.~the reference model.  The spectra, all taken at the same
time, $t = 900 \, \Jahr$, indicate how the choice of input physics
affects the results:
\begin{itemize}
\item The acceleration parameter, $\alpha_{\mathrm{acc}}$ (blue
  lines), by virtue of setting the maximum Lorentz factor of the
  electrons, strongly modifies the cutoff frequencies of both
  synchrotron and IC emission.  Changing the value by one order of
  magnitude to higher  ($\alpha_{\mathrm{acc}}
  = 1$) or lower ($\alpha_{\mathrm{acc}} = 0.01$) values reduces or
  increases the cutoff energies by approximately two orders of
  magnitude, but leaves the rising parts of the spectrum basically
  unchanged.  In particular, both spectra still go through the radio
  data points.
\item The parameter $\epsilon_e$ (green lines) setting the total
  fraction of the energy that goes into non-thermal particles is
  directly reflected in the normalisation of the spectra at all photon
  energies.  It does, on the other hand, not have any impact on the
  high-energy cutoff.  Both spectra with higher and lower values match
  the data considerably worse than our reference simulation does.
\item The number fraction of accelerated electron, $\epsilon_n$ (red
  lines), has a comparably little influence on the spectra.  The
  effect of lowering or rising its value has the opposite effect of
  lowering or rising $\epsilon_e$: distributing the same amount of
  energy among more or less accelerated electrons leads to slightly
  lower or higher spectral power, respectively.
\item The magnetisation of the medium, $\epsilon_b$ (orange lines),
  affects the spectra in a more complex way than the other three
  parameters.  The magnetic field plays a role in setting the maximum
  energy of the electrons, and, most importantly, it directly
  determines the synchrotron losses.  Stronger magnetic field
  ($\epsilon_b = 10^{-2}$) translates into a brighter synchrotron
  emission.  Consequently, the electrons cool faster, which leads to a
  much weaker IC emission, in particular at high energies.  Setting
  $\epsilon_b = 10^{-4}$ has the opposite effect: the electrons lose
  less energy due to synchrotron radiation and, therefore, emit
  stronger in the IC bands.
\end{itemize}

\section{Summary and conclusions}
\label{Sek:Sum}

Explaining the radiation observed in many galactic SNRs, characterised
by non-thermal emission across a very broad band of photon energies as
well as by a complex geometry, requires an approach combining
multi-dimensional hydrodynamical simulations and a detailed modelling
of the non-thermal radiation processes.  To this end, we have adapted
the methods used previously for computing the non-thermal emission of
relativistic outflows to the situation of multi-dimensional
non-relativistic SNRs expanding into a clumpy ISM.

A fraction of the electrons in the gas are accelerated to extremely
relativistic energies when the shock wave passes across a fluid
element and they subsequently emit synchrotron and inverse-Compton
radiation.  We apply a simple post-processing tool, \emph{SPEVita},
based on the relativistic code SPEV
\cite{Mimica_et_al__2004__aap__SyntheticX-raylightcurvesofBLLacsfromrelativistichydrodynamicsimulations}.
We follow the evolution of Lagrangian tracer particles advected with
the flow.  Upon passage of the shock wave, we set up a power-law
distribution of the electrons.  Starting from these initial states, we
evolve the distribution taking into account adiabatic
compression/expansion by the flow and energy losses due to synchrotron
and IC radiation.  We finally compute the detailed spectra of these
two radiative processes by integrating the corresponding emissivities
for all tracer particles.  From these, we can straightforwardly
obtain, e.g.~spectra or two-dimensional emission maps.

Our method is less elaborate than, e.g.~the scheme of
\cite{Ellison_et_al__2007__apj__ParticleAccelerationinSupernovaRemnantsandtheProductionofThermalandNonthermalRadiation,Patnaude_et_al__2009__apj__TheRoleofDiffusiveShockAccelerationonNonequilibriumIonizationinSupernovaRemnants,Lee_et_al__2013__apj__ACR-hydro-NEIModelofMulti-wavelengthEmissionfromtheVelaJr.SupernovaRemnant(SNRRXJ0852.0-4622)}
treating diffusive shock acceleration self-consistently.  Therefore,
as a first main step, we assessed its applicability in SNR models by
computing a small set of emission models based on the same
hydrodynamical simulation and varying some of the free parameters of
the physics of non-thermal particles and their emission.  We selected
one of the three-dimensional simulations by
\cite{Obergaulinger_et_al__2014__mnras__HydrodynamicsimulationsoftheinteractionofsupernovashockwaveswithaclumpyenvironmentthecaseoftheRXJ0852.0-4622(VelaJr)supernovaremnant},
consisting of an energetic bipolar explosion expanding into an ISM
containing several large clouds.  Although this particular
hydrodynamical model was designed as a possible scenario for the
observational appearance of SNR \VelaJr~(Vela Jr.), we do not intend
to match the observational data of this or any other individual SNR
here.  We do, however, try to roughly reproduce the main features
characteristic of most SNR.

The results of this first set of models show a reasonable quantitative
agreement with the data of the Vela Jr.~SNR.  In particular, we
clearly find the broad-band non-thermal emission with peaks in the
UV-to-X-ray and GeV-to-TeV energies, produced by synchrotron and
inverse-Compton scattering, with fluxes that are of the right order of
magnitude.  Geometrically, the emission is dominated by the immediate
post-shock region.  In addition, large clouds hit by the shock wave,
can show up prominently, in particular in soft X-ray bands, after the
shock wave has swept across them.

Instead of using different hydrodynamic models, we varied the
parameters that govern our model for the acceleration and emission of
non-thermal electrons, viz.~the acceleration efficiencies, the
relative energy and number densities of the accelerated particlces,
and the magnetisation.  Most of these changes yield a significantly
worse agreement with the observational data, which suggests that we
might use future, more detailed models to constrain their likely
values in SNRs, in particular if the remaining potential degeneracies
can be reduced by complimentary observations, e.g., of the magnetic
field in the vicinity of the explosion.

Having found a qualitative agreement between models and general
observational features of SNRs, we plan to extend the present analysis
by considering a wider range of physical parameters for the
acceleration and emission processes and apply it to a larger set of
(magneto-)hydrodynamical simulations in order to investigate the
observational consequences of different sources for asymmetries in
SNRs.

\section{Acknowledgements}
\label{Sek:Ackno}

We thank Ewald M{\"u}ller for stimulating discussions.  AFI was
partially supported through the Grant of RF ``11.G34.31.0076''.  MO,
PM, and MAA acknowledge support from the European Research Council
(grant CAMAP-259276), and from the Spanish Ministerio de Ciencia e
Innovaci{\'o}n (grant  AYA2013-40979-P \emph{Astrof{\'i}sica
  Relativista Computacional}) and from the Valencian Conselleria
d'Educaci{\`o} (PROMETEO-2009-103).

\bibliographystyle{elsarticle-harv}

\end{document}